\documentclass[conference]{IEEEtran}
\IEEEoverridecommandlockouts
\usepackage{cite}
\usepackage{amsmath,amssymb,amsfonts}

\usepackage{multirow}
\usepackage{booktabs}
\usepackage[ruled,linesnumbered]{algorithm2e}
\usepackage[normalem]{ulem}
\usepackage{balance}
\usepackage{threeparttable}
\usepackage{url}

\usepackage{graphicx}
\usepackage{textcomp}
\usepackage{xcolor}
\usepackage{hyperref}
\usepackage{subcaption}

\def\BibTeX{{\rm B\kern-.05em{\sc i\kern-.025em b}\kern-.08em
    T\kern-.1667em\lower.7ex\hbox{E}\kern-.125emX}}

\begin{document}

\title{Logic Optimization Meets SAT: A Novel Framework for Circuit-SAT Solving}

\author{
	\IEEEauthorblockN{
	Zhengyuan Shi$^{1,4}$,
        Tiebing Tang$^1$, 
        Jiaying Zhu$^{1,4}$, 
        Sadaf Khan$^1$, \\
        Hui-Ling Zhen$^2$, 
        Mingxuan Yuan$^2$, 
        Zhufei Chu$^3$ 
        and Qiang Xu$^{1,4}$\thanks{Corresponding author: Qiang Xu (qxu@cse.cuhk.edu.hk)}
        } 

\IEEEauthorblockA{$^1$\textit{Department of Computer Science and Engineering}, \textit{The Chinese University of Hong Kong}, Hong Kong S.A.R.\\}
\IEEEauthorblockA{$^2$\textit{Noah's Ark Lab}, \textit{Huawei}, Hong Kong S.A.R.\\}
\IEEEauthorblockA{$^3$\textit{Faculty of Electrical Engineering and Computer Science}, \textit{Ningbo University}, Ningbo, China \\}
\IEEEauthorblockA{$^4$\textit{National Center of Technology Innovation for EDA}, Nanjing, China \\}
} 

\maketitle

\begin{abstract}


The Circuit Satisfiability (CSAT) problem, a variant of the Boolean Satisfiability (SAT) problem, plays a critical role in integrated circuit design and verification. However, existing SAT solvers, optimized for Conjunctive Normal Form (CNF), often struggle with the intrinsic complexity of circuit structures when directly applied to CSAT instances. To address this challenge, we propose a novel preprocessing framework that leverages advanced logic synthesis techniques and a reinforcement learning (RL) agent to optimize CSAT problem instances. The framework introduces a cost-customized Look-Up Table (LUT) mapping strategy that prioritizes solving efficiency, effectively transforming circuits into simplified forms tailored for SAT solvers. Our method achieves significant runtime reductions across diverse industrial-scale CSAT benchmarks, seamlessly integrating with state-of-the-art SAT solvers. Extensive experimental evaluations demonstrate up to 63\% reduction in solving time compared to conventional approaches, highlighting the potential of EDA-driven innovations to advance SAT-solving capabilities.

\end{abstract}


\section{Introduction} \label{Sec:Intro}


The Circuit Satisfiability (CSAT) problem lies at the heart of many critical tasks in Electronic Design Automation (EDA), such as logic equivalence checking (LEC), automatic test pattern generation (ATPG), and logic synthesis (LS). It involves determining whether a Boolean circuit has an input assignment that satisfies its primary output. As a specialized variant of the Boolean Satisfiability (SAT) problem, CSAT is known to be NP-complete~\cite{cook1971complexity}, making it both theoretically challenging and practically important in integrated circuit (IC) design and verification.


Modern SAT solvers, such as MiniSAT, Kissat, and CaDiCaL~\cite{sorensson2005minisat, fleury2020cadical}, have been highly effective in solving problems represented in Conjunctive Normal Form (CNF). However, when applied to CSAT problems, these solvers face inefficiencies stemming from the required transformation of circuits into CNF format. This transformation often disrupts the natural structure of circuits, leading to representations that are less solver-friendly and more complex to process~\cite{ostrowski2002recovering,een2005effective}. As a result, solving CSAT instances may become unnecessarily time-consuming, particularly for large-scale, industrial circuits.



While preprocessing techniques for CNF have been extensively explored~\cite{ostrowski2002recovering, een2005effective, subbarayan2005niver, condrat2007grobner}, similar advancements tailored specifically for circuit-based representations remain underdeveloped. Circuit-specific methods often focus on reducing circuit size or area~\cite{een2007applying}, which, while useful for design optimization, do not directly address the complexity of SAT solving. This gap highlights the need for a new approach that bridges the structural strengths of circuit representations with the solving power of modern SAT solvers.


To address these limitations, we propose a novel preprocessing framework tailored to CSAT problems. Our approach integrates EDA techniques with reinforcement learning (RL) to transform CSAT instances into SAT-friendly formats. By leveraging logic synthesis tools and a cost-customized Look-Up Table (LUT) mapping strategy, our framework achieves significant reductions in solving complexity. Unlike traditional preprocessing methods, our approach explicitly prioritizes SAT-solving efficiency rather than widely-used optimization metrics like area or delay. Specifically, the main contributions of our work include:

\begin{itemize}
    \item \textbf{EDA-Driven Preprocessing Framework}: We introduce a comprehensive preprocessing pipeline that combines logic synthesis techniques with reinforcement learning to optimize CSAT instances for efficient SAT solving. This integration bridges the gap between raw circuit structures and modern SAT solvers.
    \item \textbf{RL-Guided Logic Synthesis for SAT Solving}: We formulate logic synthesis as a Markov Decision Process (MDP) and train an RL agent to identify synthesis sequences that minimize solving complexity. The agent leverages circuit features inspired by~\cite{hosny2020drills,zhu2020exploring} and structural circuit embeddings from DeepGate2~\cite{shi2023deepgate2}.
    \item \textbf{Cost-Customized LUT Mapping for SAT Solving}: We develop a novel LUT mapping strategy that incorporates a customized cost metric reflecting branching complexity during SAT solving. This approach diverges from conventional LUT mapping methods~\cite{francis1991chortle,soeken2018epfl} by focusing on solving efficiency.
\end{itemize}

Using industrial-scale benchmarks derived from LEC and ATPG tasks, we demonstrate significant runtime reductions up to 63\% compared to baseline approaches. Our method integrates seamlessly with state-of-the-art SAT solvers, such as Kissat and CaDiCaL.


\section{Related Work} \label{Sec:Related}

\begin{figure*}
    \centering
    \includegraphics[width=0.85\linewidth]{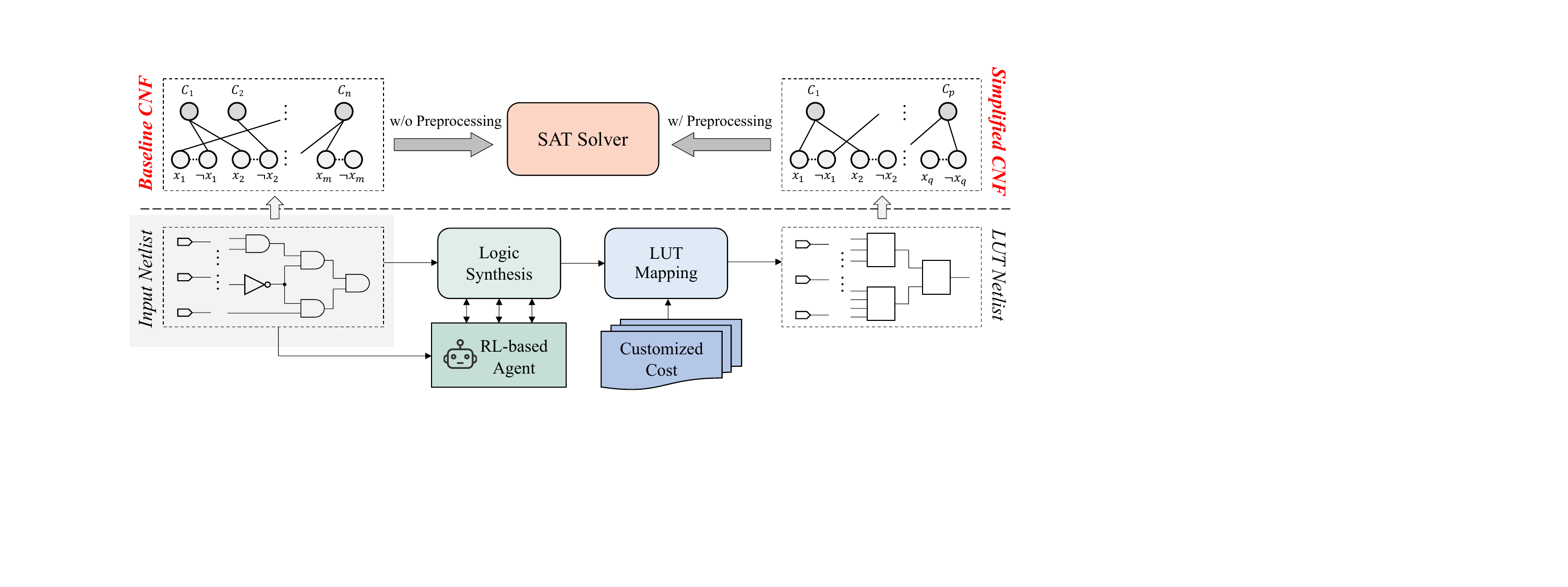}
    \caption{Overview of EDA-Driven SAT preprocessing framework}
    \label{fig:overview}
    \vspace{-10pt}
\end{figure*}

\subsection{SAT and Circuit SAT Problems}
SAT problem holds a pivotal position in hardware design, software verification, planning and scheduling, which seeks to identify if there exists at least one assignment that makes a given Boolean formula True. The modern SAT solvers, central in addressing SAT problems, predominantly rely on the Conflict-Driven Clause Learning (CDCL) algorithm~\cite{marques2021conflict}. Within the backbone CDCL algorithm, CNF has become a standard format in SAT solving, capable of representing arbitrary Boolean formulas. It is structured as a conjunction of clauses, denoted by $\phi = (C_1 \land C_2 \land \ldots)$, with each clause being a disjunction of variables or their negations, e.g., $C_i = (x_1 \lor \neg x_2 \lor \ldots)$. Over recent decades, numerous CNF-based heuristics have been developed to enhance SAT solving efficiency, including advanced clause management strategies~\cite{audemard2009predicting} and innovative branching heuristics~\cite{moskewicz2001chaff, shi2022satformer}. 

Typically, practical CNF-modeled problems often result in redundant clauses and increased complexity. For example, a tournament scheduling problem with $n$ teams can escalate to $n^6$ clauses~\cite{bejar2000solving}. Consequently, problem reformulation, as known as preprocessing, is essential before solving. Efforts like~\cite{een2005effective, subbarayan2005niver, ostrowski2002recovering} target minimizing the CNF instances by resolution and graph transformation, resulting in faster SAT solving. 

One of the most representative SAT variants in the realm of EDA is the Circuit SAT (CSAT) problem. In contrast to the general SAT problem, the CSAT problem determines the satisfiability of a Boolean circuit. Over the past decades, circuit-based SAT solvers have been developed to tackle the CSAT problems~\cite{wu2007qutesat, zhang2021circuit, li2022deepsat}. Despite these solvers conducting logic reasoning in natural formats, they still exhibit lower performance levels compared to CNF-based SAT solvers due to the inefficient circuit-based preprocessing and heuristics. For example, the previous circuit-based preprocessing approaches, such as ~\cite{een2007applying}, cannot promise to reduce solving time as they only scale down circuits. In this work, we introduce an efficient and effective circuit-based preprocessing framework to directly mitigate solving complexity by integrating EDA tools and reinforcement learning guidance. 

\subsection{Boolean Circuit Optimization}
Boolean circuits, representable as directed acyclic graphs (DAGs) $\mathcal{G}(\mathcal{V}, \mathcal{E})$ with logic gates and wires serving as nodes $\mathcal{V}$ and edges $\mathcal{E}$, respectively, are pivotal in computational logic. In the EDA community, a wealth of efficient algorithms has been developed for circuit optimization. Logic Synthesis (LS) is a prime example, involving DAG-based transformation techniques to modify circuit's local structure for the optimization of performance, power, and area (PPA). 

Besides, technology mapping, a key component in EDA flow, fine-tunes performance metrics under specific constraints~\cite{berkelaar1988technology} and maps the intermediate circuits into standard cells pre-defined in technology library. One widely used technique for technology mapping is Look-Up Table (LUT) mapping~\cite{francis1991chortle}. LUT mapping involves the transformation of gate combinations into $k$-LUTs, where each $k$-LUT represents a configurable memory unit capable of implementing arbitrary Boolean functions with $k$ variables. 

The advent of circuit representation learning~\cite{chen2024large} marks a progressive trend in the field, focusing on deriving neural representations of circuits to facilitate downstream tasks. The DeepGate family~\cite{li2022deepgate, shi2023deepgate2, shi2024deepgate3}, for instance, integrates structural and functional aspects of circuits and has shown impressive results in testability analysis~\cite{shi2022deeptpi} and SAT solving~\cite{li2022deepsat}. In our research, we tailor existing EDA tools for specific requirements, utilizing DeepGate2's~\cite{shi2023deepgate2} circuit knowledge learning capabilities to inform our circuit optimization decisions.
\section{Methodology} \label{Sec:Method}

\subsection{Overview of the Proposed Framework}

Fig.~\ref{fig:overview} shows the overview of the proposed circuit-based preprocessing framework. 
Given a CSAT problem represented in And-Inverter Graphs (AIGs), which only contain three gate types (primary input, AND and NOT) by~\cite{biere2006aiger}, the baseline pipeline converts it into CNF. 

In our proposed framework, we first apply a predetermined sequence of AIG transformation operations to unify the distribution of input circuits~\cite{li2022deepsat}. Then, we utilize logic synthesis technique to reformulate SAT instance. Since the logic synthesis process can be considered as a sequential decision-making task, we train an agent to minimize the solving complexity by reinforcement learning (RL). In the RL framework, the state comprises embeddings of the initial AIG netlist and features of intermediate circuits. Additionally, the RL agent operates within a discrete action space, where each action corresponds to executing a synthesis operation. We approximate the complexity of SAT solving based on the number of variable branching times and define the reduction in branching times as the reward for RL agent. 

Finally, we employ the LUT mapping technique to hide the internal logic of circuits. To further decrease the solving time, we design a cost-customized mapping operation with the objective of minimizing the branching times during SAT solving, rather than solely focusing on reducing the problem size. After the LUT mapping, we obtain the LUT netlist with a relatively small number of cells. To ensure compatibility with modern SAT solvers, we transform the LUT netlist back into a simplified CNF. 

\subsection{Logic Synthesis Recipe Exploration} \label{Sec:Syn}
We formulate LS as a Markov Decision Process (MDP) process, with the optimization objective of reducing the solving time for a given input instance. Consequently, such sequential decision-making process entails iteratively selecting a LS operation as an action according to the current circuit state. 

\subsubsection{Reinforcement Learning Formulation}
To explore the optimal LS sequence (i.e. LS recipe), we employ Deep Q-learning algorithm in this section to train a reinforcement learning (RL) agent. The RL agent is trained to make informed decisions to minimize solving time by learning from its interactions with the circuit environment. 

\begin{figure}[!t]
    \centering
    \includegraphics[width=0.9\linewidth]{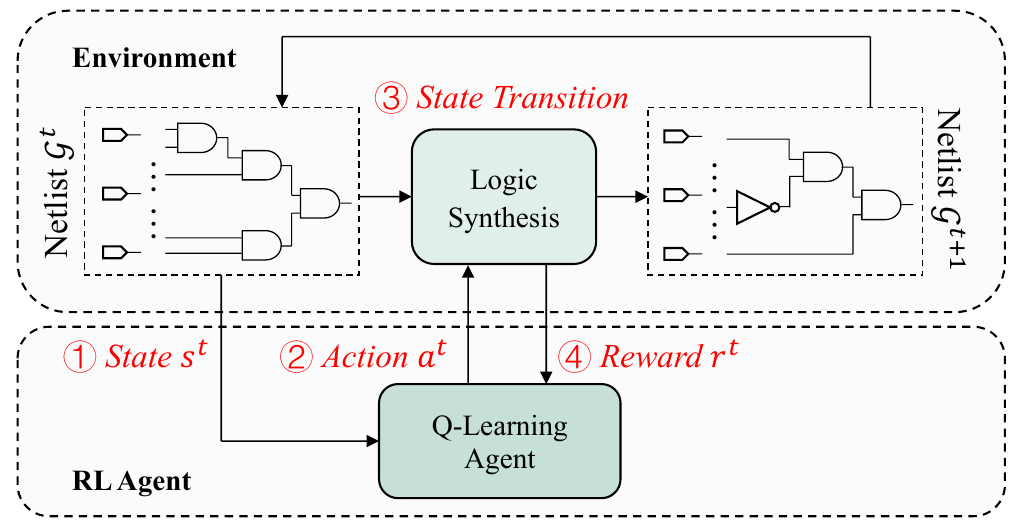}
    \caption{RL Agent-Environment interaction}
    \label{fig:rl}
    \vspace{-10pt}
\end{figure}

As shown in Fig.~\ref{fig:rl}, at each step $t$ (where $t = 0, 1, ..., T-1$), the features of the current netlist $\mathcal{G}^t$ are extracted to form the state $s^t$. Subsequently, the Q-learning agent utilizes $s^t$ as input and selects an action $a^t$. Then, the environment including the LS tool performs the synthesis operation based on $a^t$ and transforms the netlist $\mathcal{G}^t$ into a new netlist $\mathcal{G}^{t+1}$. At the end of this step, the environment provides RL agent with a reward $r^t$. The above process is repeated for subsequent steps until $t = T-1$. 
During the training process, the objective of RL agent is to maximize the cumulative sum of the reward values. As a result, the trained RL agent is able to select LS operations step by step to achieve the optimal solving time reduction. 

\subsubsection{State}
Inspired by~\cite{hosny2020drills, zhu2020exploring}, we extract the following representative features $\mathcal{E}(\mathcal{G}^t)$ of the netlist $\mathcal{G}^t$ into state $s^t$: 
\begin{itemize}
    \item Ratio between the area of netlist $\mathcal{G}^t$ and netlist $\mathcal{G}^0$.
    \item Ratio between the depth of netlist $\mathcal{G}^t$ and netlist $\mathcal{G}^0$. 
    \item Ratio between the wire counts of netlist $\mathcal{G}^t$ and netlist $\mathcal{G}^0$.
    \item Proportion of AND gates in the total gates of netlist $\mathcal{G}^t$.
    \item Proportion of NOT gates in the total gates of netlist $\mathcal{G}^t$.
    \item Average balance ratio of netlist $\mathcal{G}^t$. We formally define the average balance ratio $br$ in Eq.~\eqref{Eq:br}, where $\mathcal{P}_{i1}$, $\mathcal{P}_{i2}$ are predecessors of two fanin AND gate $i$, and $d$ denotes the depth.
\end{itemize}
\begin{equation} \label{Eq:br}
    br = \sum_{i \in AND}\frac{abs(d_{\mathcal{P}_{i1}}-d_{\mathcal{P}_{i2}})}{max(d_{\mathcal{P}_{i1}}, d_{\mathcal{P}_{i2}})}\ /\ \# AND
\end{equation}

The above features exhibit characteristics that can be effectively optimized by various logic synthesis operations. Consequently, the RL agent can intelligently select appropriate operations based on these features. For example, when a netlist shows a higher average balance ratio, indicating an imbalanced structure in DAG, the agent is more inclined to select the action of operation \textit{balance}. Additionally, a large area ratio prompts the agent to opt for operations that aim to minimize the circuit area, such as \textit{rewrite}. 

Besides, the state $s^t$ is incorporated with the primary output embeddings of the initial netlist obtained by pretrained DeepGate2~\cite{shi2023deepgate2} (denoted as $\mathcal{D}(\mathcal{G}^0)$), which contains rich structural and functional information of the initial problem instance. 
\begin{equation} \label{Eq:state}
    s^t = \text{Concatenate}(\mathcal{E}(\mathcal{G}^t), \mathcal{D}(\mathcal{G}^0))
\end{equation}

\subsubsection{Action}
The action space $\mathcal{A}$ of agent is discrete and encompasses LS operations. In this paper, we set the available operations as follows: \textit{rewrite}~\cite{mishchenko2006dag}, \textit{refactor}~\cite{brayton1982decomposition}, \textit{balance}~\cite{cortadella2003timing}, \textit{resub}~\cite{sato1991boolean} and \textit{end} (marking the end of the LS process).  
The selection of these operations does not imply that the proposed framework lacks support for additional synthesis operations beyond this set. Instead, these operations are chosen due to their widespread usage and commonality. If there is a need to include additional optional operations, it can be accomplished by expanding the action space within the framework. 

\subsubsection{State Transition}
The state transition function is implemented by the LS tool. We call the tool with a specific LS operation determined by RL agent to transform the current netlist $\mathcal{G}^t$ to another functional-equivalent but simplified netlist. More formally, we note the given netlist as $\mathcal{G}^t$ and RL action as $a^t$ at step $t$. The state transition function is denoted as: $\mathcal{G}^{t+1} = \mathcal{F}(\mathcal{G}^t, a^t)$, where $\mathcal{G}^{t+1}$ is the updated netlist by tool. The updated netlist, $\mathcal{G}^{t+1}$, is then considered as the input for the subsequent decision step at time $t+1$. 

\subsubsection{Reward}
We consider the reduction of variable branching times during SAT solving as the reward. The reward function is shown as Eq.~\eqref{Eq:reward}, where $\Delta \#Branching$ is the difference in branching times between the final instance and the initial instance during solving. We opt for the terminated reward, which has only one non-zero reward value at the terminated step, i.e. after performing the entire sequence of synthesis operations. 
\begin{equation} \label{Eq:reward}
    r^t =  \left\{
        \begin{aligned}
            & 0, t = 0,\cdots,T-2, \\
            & \textnormal{-} \Delta \#Branching,\ t=T-1\ or\ a^t = \textit{end}
        \end{aligned}
            \right.
\end{equation}

Prior to solving the SAT instance in circuit format, we employ the LUT mapping to convert the AIG netlist into an LUT netlist. Subsequently, the transformed netlist is converted into CNF, which will be elaborated in the following section. Additionally, we choose not to minimize the solver runtime based on two primary reasons. Firstly, determining the solver runtime for hard SAT instances is time-consuming. Secondly, the solver runtime is challenging to calculate accurately due to the fluctuating CPU status in the case of easy SAT instances. We utilize variable branching times to approximate solving time. This approach allows us to include easy cases with short solver runtime in the training dataset, thereby enhancing the efficiency of training the RL agent.

\subsubsection{Policy}
In the Q-learning framework~\cite{mnih2015human}, the policy for maximizing the total reward entails selecting the action with the highest expected value. The action value function $Q$ uses the multilayer perceptron (MLP) to estimate the sum of future rewards after taking a certain action in the current state. The policy $\pi(s^t)$ and action-value function $Q_{\theta}(s^t, a)$ are shown in Eq.~\eqref{Eq:policy}, where the $\theta$ parametrizes the function $Q$. 
\begin{equation} \label{Eq:policy}
    \begin{split}
        \pi(s^t) & = \arg \max_{a, a \in \mathcal{A}}(Q(s^t, a)) \\
        Q_{\theta}(s^t, a) &= Index(MLP(s^t), a)
    \end{split}
\end{equation}

We create the target function $Q_{\hat{\theta}}(s^t, a)$ to stabilize the training process. The parameter $\hat{\theta}$ is copied from $\theta$ after $k$ steps. The parameter $\theta$ of action-value function $Q$ is updated iteratively by loss function in Eq.~\eqref{Eq:loss}, where $\gamma$ is a hyperparameter and represents discount factor. 
\begin{equation} \label{Eq:loss}
    l = ||Q_{\theta}(s^t, a^t) - r^t - \gamma \arg \max_{a \in \mathcal{A}}Q_{\hat{\theta}}(s^{t+1}, a)||_2
\end{equation}

\subsection{Cost-Customized LUT Mapping} \label{Sec:Map}
We present a cost-customized LUT mapping approach that not only hides the internal nodes during solving, but also forms an easy-to-solve instance. Unlike conventional mapping tools that focus on constructing circuits with low area or delay, our approach aims to create circuits that are easier to solve. Specifically, we begin by defining a cost metric to quantify the branching complexity of SAT solving and compute the cost values for all $4$-input LUTs. Next, we modify the cost function in the existing mapping tool to enable minimizing the overall branching complexity. Lastly, we apply the cost-customized mapper to map AIG netlist to LUT netlist and convert the LUT netlist into CNF.

\subsubsection{Branching Complexity of LUT}
\begin{figure}[!t]
    \centering
    \includegraphics[width=0.8\linewidth]{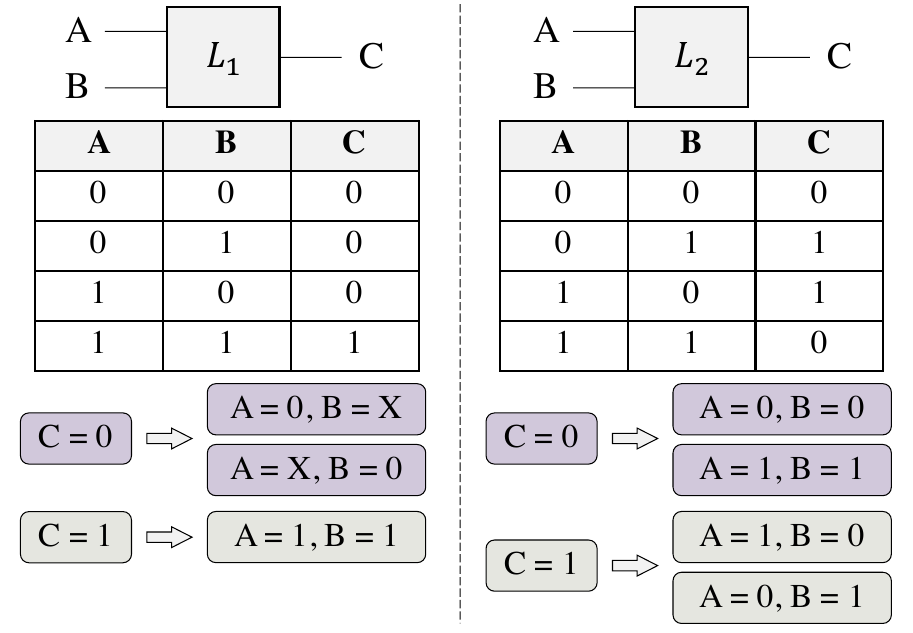}
    \caption{LUTs with different branching complexity}
    \label{fig:lut}
    \vspace{-10pt}
\end{figure}

The SAT solving process involves iteratively making variable branching decisions, where the solver selects variable assignments as either \textit{true} or \textit{false}. We formulate the solving process within the circuit domain. Specifically, given the logic value of a gate, the SAT solver makes decisions by selecting assignments for its fanin gates. We consider two 2-fanin LUTs, denoted as $L_1$ (representing an AND gate) and $L_2$ (representing an XOR gate), and list the truth tables for these LUTs in Fig.~\ref{fig:lut}.

We observe that both $L_1$ and $L_2$ have two possible combinations when the fanout pin (C) is logic-0. However, when the fanout pin is logic-1, $L_2$ has two possible combinations while $L_1$ has only one unique branch. We define the \textit{branching complexity} $C$ as the total number of possible fanin combinations. In this example, the LUT $L_1$ has $C_{L_1}=3$ and LUT $L_2$ has $C_{L_2}=4$. Therefore, $L_2$ is more hard to be solved than $L_1$. 

\begin{algorithm}[!t]
\caption{CSAT preprocessing framework}
\label{Algo:Overview}
\KwIn{The input circuit instance $\mathcal{G}_{in}$}
\KwOut{The output simplified CNF instance $\phi_{out}$}
    \tcc{Initialize circuit graph $\mathcal{G}^0$ in AIG}
    \eIf{input instance is in AIG}{
      $\mathcal{G}^0 = \mathcal{G}_{in}$\;
      }{
      $\mathcal{G}^0 = \textit{aigmap}(\mathcal{G}_{in})$\;
      }
      
    \tcc{Explore logic synthesis recipe by RL-agent (see Section~\ref{Sec:Syn})}
    Define maximum step count: $T$\;
    $t = 0$\;
    \While{$t < T$} {
        \tcc{Prepare the state of current circuit}
        $s^t = \text{Concatenate}(\mathcal{E}(\mathcal{G}^t), \mathcal{D}(\mathcal{G}^0))$\;
        \tcc{Make decision to choose a logic synthesis operation}
        $a^t = \pi(s^t) = \arg \max_{a, a \in \mathcal{A}}(Q(s^t, a))$\;
        \If{$a^t$ is \textit{end}} {
            break\;
        }
        \tcc{Perform logic synthesis for state transition}
        $\mathcal{G}^{t+1} = \mathcal{F}(\mathcal{G}^t, a^t)$\;
        $t = t+1$\;
    }
    
    \tcc{Perform cost-customized LUT mapping (see Section~\ref{Sec:Map})}
    Define LUT costs: $C$ \;
    $\mathcal{G}_{LUT} = \textit{mapper}(\mathcal{G}^t, C)$\;
    \tcc{Convert to CNF}
    $\phi_{out} = \textit{lut2cnf}(\mathcal{G}_{LUT})$\;
    \Return $\phi_{out}$\;
\end{algorithm}

\subsubsection{Cost-Customized Mapper}
Since a large number of possible branches result in an exponential search space, various heuristics~\cite{liang2016learning, marques2021conflict} are employed to reduce the number of potential branches and improve the efficiency of the solving process. Besides, utilizing techinque mapping tools facilitates the concealment of intermediate gates and decreases the overall number of branches~\cite{een2007applying}. 

In contrast to~\cite{een2007applying} that minimizes the size of the post-mapping netlist, our approach prioritizes mapping the logic into cells with lower branching complexity. Our approach is motivated by the observations that instances containing a significant proportion of XOR gates often require more time to be solved~\cite{haanpaa2006hard, dudek2017hard}, where the branching complexity of XOR gate ($L_2$ in Fig.~\ref{fig:lut}) is higher than other gates.

To prioritize the LUTs during mapping process, we employ a strategy to enumerate all $4$-LUTs and integrate their branching complexity into the cost function. We implement a cost-customized mapper based on mockturtle~\cite{soeken2018epfl} by modifying the area cost to reflect the branching complexity of each LUT and fixing the delay cost as a constraint. By executing a sequence of operations aimed at minimizing the total cost, the resulting post-mapping netlists exhibit minimal overall branching complexity.Our preprocessing strategy ends with a LUT to CNF transformation~\cite{ling2005fpga} for compatibility. 

As a result, we formally depict our overall circuit-based preprocessing framework coupled with RL-based logic synthesis exploration and cost-customized LUT mapping in Algorithm~\ref{Algo:Overview}. The framework begins by transforming the input circuit instance $\mathcal{G}_{in}$ into AIG format, using $aigmap$ technique in ABC~\cite{mishchenko2007abc}. Then, the RL agent (elaborated in Section~\ref{Sec:Syn}) proceeds to select logic synthesis operations step by step through making action $a^t$. During this process (line 6-16), the circuit is synthesized multiple times to create a simplified circuit. Next, the framework continues to perform cost-customized LUT mapping on the AIG netlist $\mathcal{G}^t$, resulting in a compact LUT netlist $\mathcal{G}$. Finally, the LUT netlist is encoded into CNF as output $\phi_{out}$. Due to these simplification approaches, $\phi_{out}$ has a lower solving complexity than $\phi_{in}$ and the subsequent SAT solving procedures can be more efficient. We will assess the effectiveness of our preprocessing by the following experiments.

\section{Experiments} \label{Sec:Experiment}

\begin{figure*}[!t]
    \centering
    \subfloat{
    \begin{minipage}[t]{0.42\linewidth}
        \centering
        \includegraphics[width=1.0\textwidth]{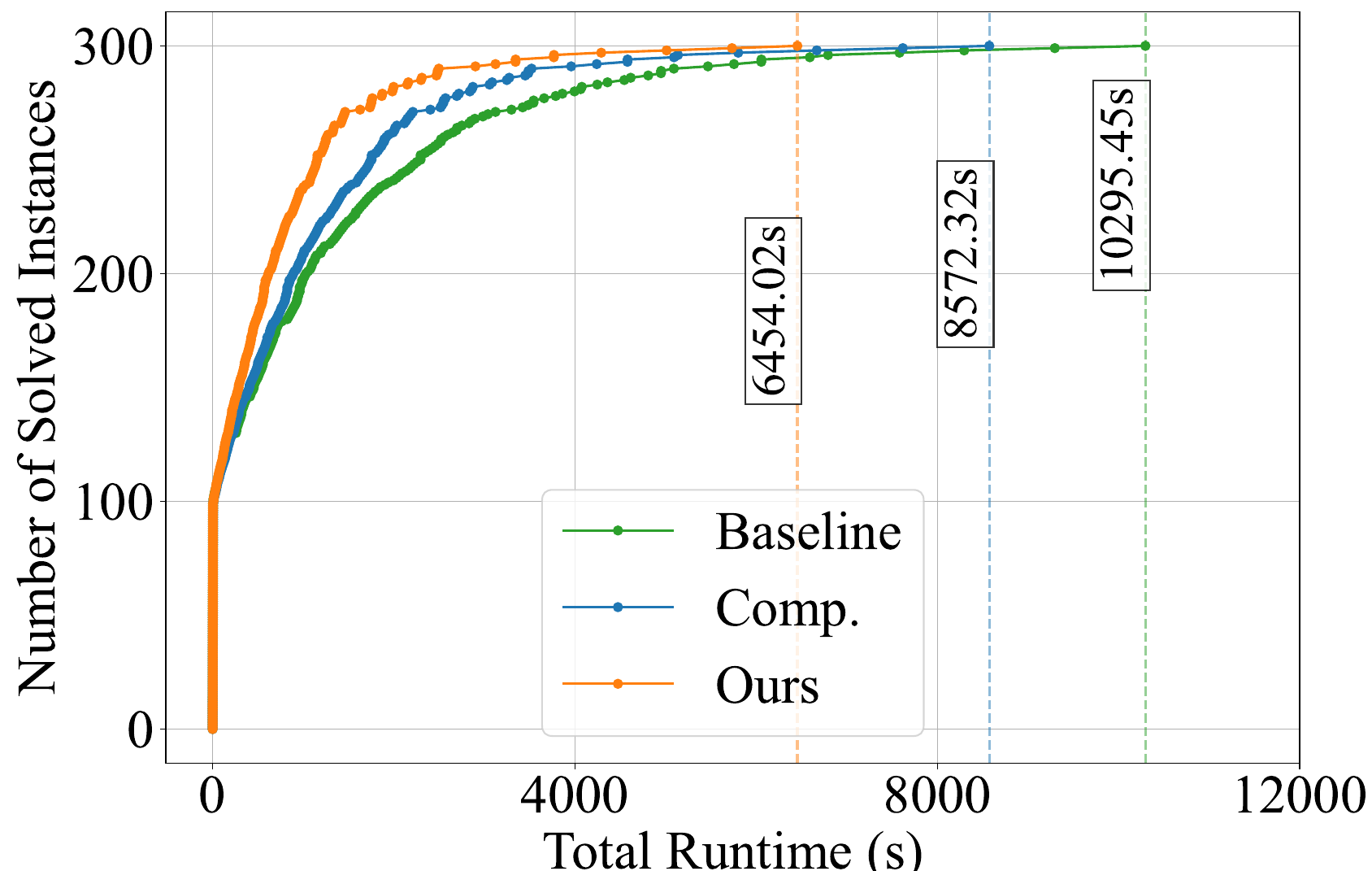}
        \subcaption{Kissat}
    \end{minipage}
    }
    \subfloat{
    \begin{minipage}[t]{0.42\linewidth}
        \centering
        \includegraphics[width=1.0\textwidth]{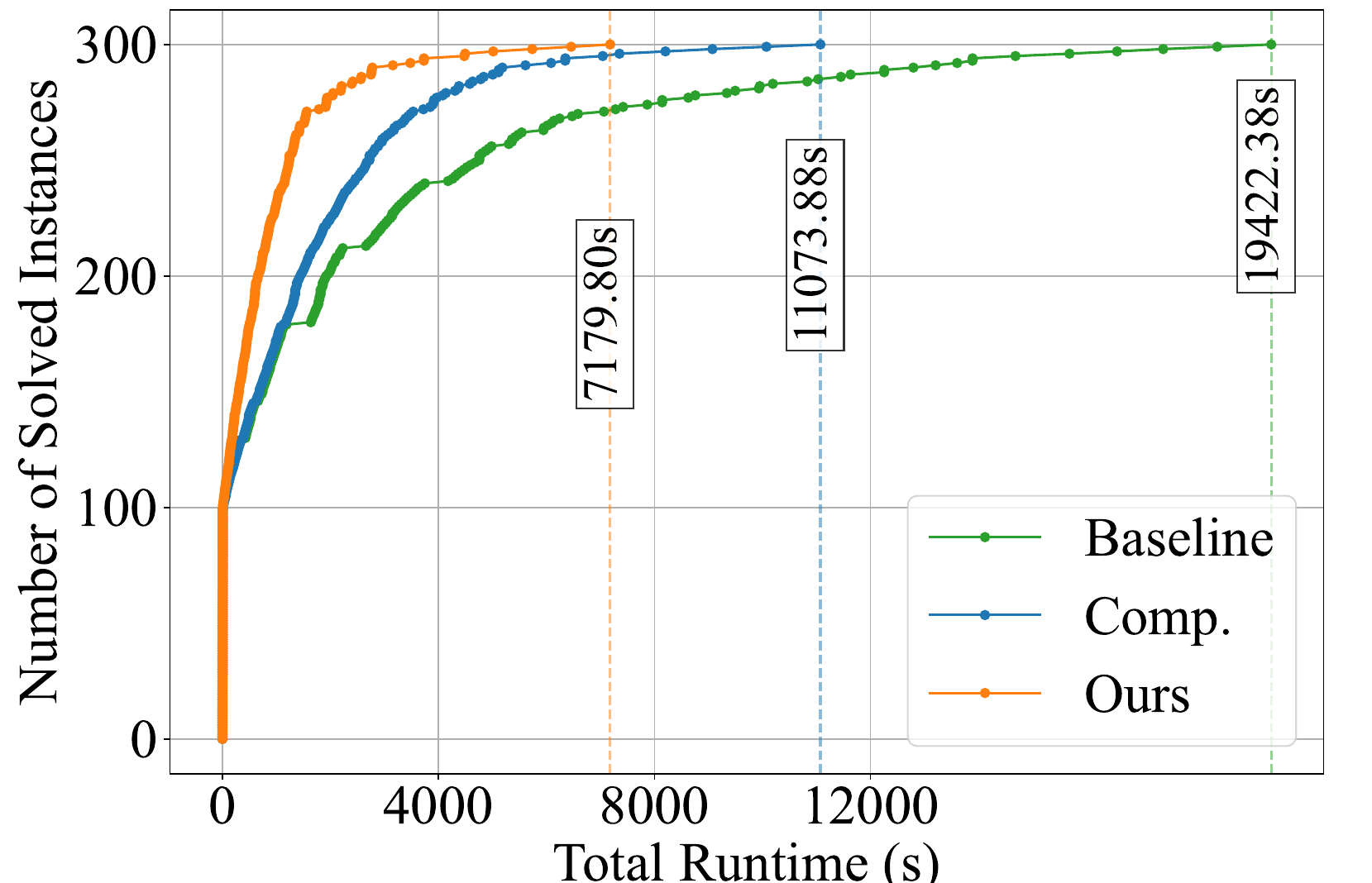}
        \subcaption{Cadical}
    \end{minipage}
    }
    \caption{Runtime Comparison with Baseline and Comp.}
    \label{FIG:Res}
    \vspace{-10pt}
\end{figure*}

\subsection{Experiment Settings}

\begin{table}[!t]
\centering
\caption{Statistics of training dataset} \label{TAB:dataset}
\begin{tabular}{@{}lllll@{}}
\toprule
           & Avg.      & Std.      & Min. & Max.   \\ \midrule
\# Gates   & 4,299.06  & 4,328.16  & 60   & 24,178 \\
\# PIs     & 43.66     & 25.17     & 6    & 102    \\
Depth      & 66.43     & 19.98     & 18   & 138    \\
\# Clauses & 10,687.28 & 10,801.96 & 131  & 60,294 \\ 
Time (s)   & 2.01      & 1.96      & 0.04 & 6.68   \\ \bottomrule
\end{tabular}
\vspace{-5pt}
\end{table}

We collect $200$ easy (solving time $0.04s-6.68s$) instances into the RL training dataset and $300$ hard instances for testing. All instances are in AIG format and derived from both industrial logic equivalence checking (LEC) and automatic test pattern generation (ATPG) problems. For LEC instances, we modify original industrial datapath circuits and connect their primary outputs (POs) through XOR gates to form CSAT instances, where satisfiability indicates non-equivalent functionality between circuits. For ATPG instances, we introduce stuck-at faults into industrial circuits and connect the POs of faulty and fault-free circuits through XOR gates, where satisfiable assignments serve as test patterns for fault detection. We construct $200$ LEC instances and $100$ ATPG instances in total, where 200 easy instances are selected for RL training based on their clause count after CNF transformation and solving time without preprocessing. The statistics of training instances are shown in Table~\ref{TAB:dataset}, where \# Clauses shows the number of clauses after transforming AIG instances into CNFs and Time represents the solving time without any preprocessing.

In the RL setting, we assign the maximum number of steps per episode as $T=10$ and the discount factor $\gamma=0.98$. During each training episode, the agent randomly selects an instance and makes decisions aimed at transforming the instance to minimize the number of branching times in SAT solving. We conduct a total of $10,000$ episodes, i.e., the RL agent preprocesses instances for $10,000$ times during RL training. The batch size is $32$ and learning rate is $10^{-5}$. 

We opt for the ABC tool~\cite{mishchenko2007abc} to synthesize the circuit and build a cost-customized mapper based on mockturtle~\cite{soeken2018epfl} to map the circuit in LUT netlists. The solver runtime limitation is $1,000$s. Any cases that exceed this time limit are marked with TO (Time-out) and are considered to have a solving time of $1,000$ seconds ($\mathcal{T}_{solve}=1,000 s$). Each instance solving is conducted on a single core of Intel Xeon CPU E5-2630. 

\subsection{Runtime Comparison}
Figure~\ref{FIG:Res} illustrates the relationship between the number of instances solved and overall runtime (including RL model inference time, transformation time and solving time). The Baseline setting represents the conventional solving pipeline, encoding the circuit-based instances directly into CNFs. Our experiment evaluates the effectiveness of our proposed framework using both Kissat solver (version 4.0.0~\cite{heule2024proceedings}) and CaDiCaL solver (version 2.0.0~\cite{biere2024cadical}), with our specific settings distinguished by the orange line (denoted as Ours).

To showcase the efficiency of our preprocessing framework, we compare the solving time with another circuit-based approach\footnote{We do not include comparisons with CNF-based preprocessing approaches because our framework is not mutually exclusive with them and can be used in conjunction. Pratically, we keep the default CNF-based preprocessing.}~\cite{een2007applying}, denoted as Comp. . To the best of our knowledge, although such approach has been around for over a decade, this is the only circuit-based SAT preprocessing technique. 

According to Figure~\ref{FIG:Res}, our proposed circuit-based proprocessing framework can significantly reduce runtime. To be specific, solving these 300 instances necessitates 19,422.38 seconds for the CaDiCaL Baseline pipeline and 11,073.88 seconds for the comparative pipeline (Comp.). In contrast, our method completes the task in a total of 7,179.80 seconds, marking a reduction of $63.03\%$ and $35.16\%$ in solving time compared to Baseline and Comp., respectively. 

\subsection{Ablation Studies}
In the following experiments, we investigate the effectiveness of our RL-based agent and cost-customized mapper. We keep the same experiment environments and test cases above. 

\begin{figure} [!t]
    \centering
    \includegraphics[width=0.8\linewidth]{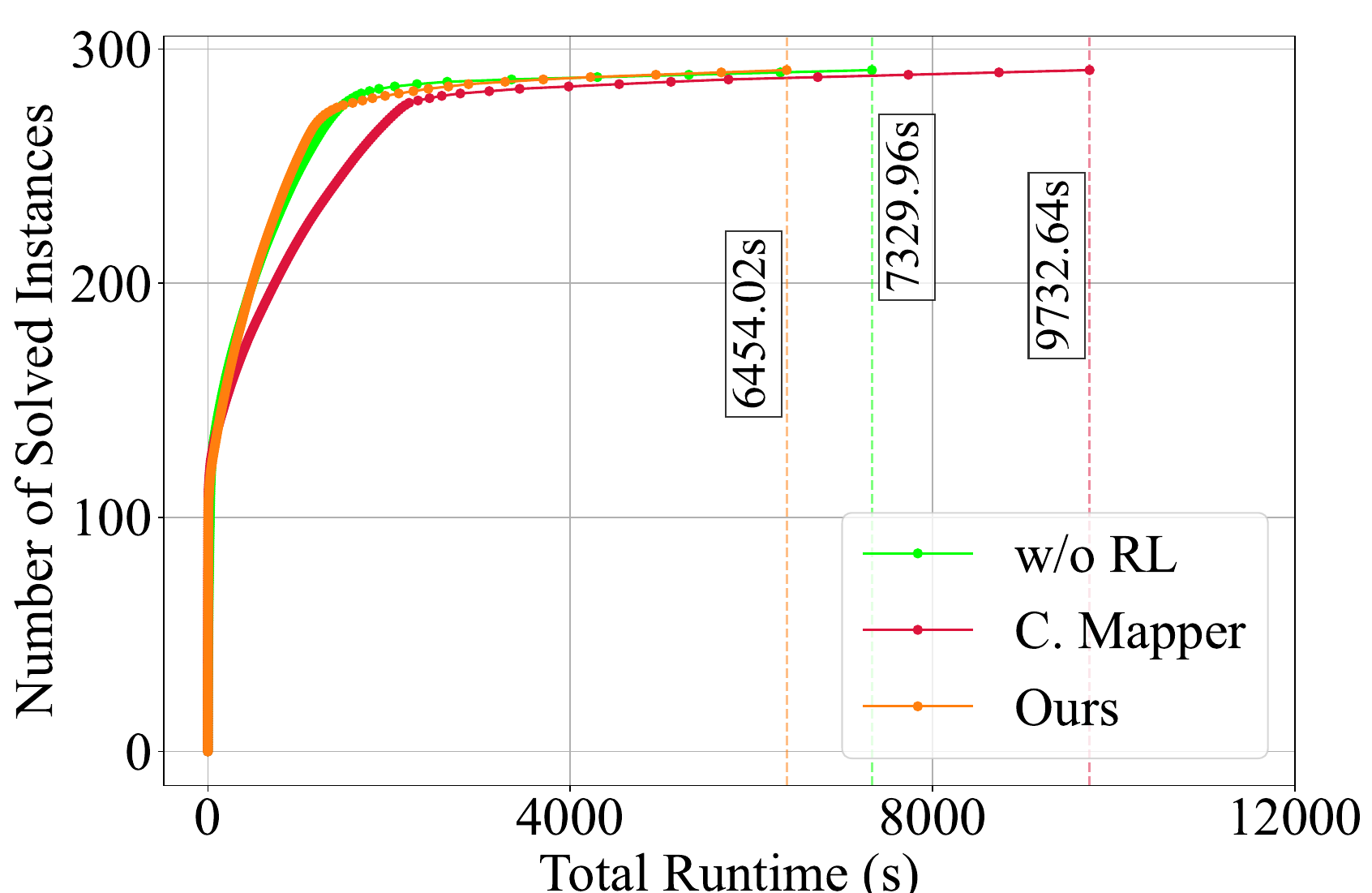}
    \caption{Ablation Study}
    \vspace{-10pt}
    \label{fig:abs}
\end{figure}

\subsubsection{Effectiveness of Logic Synthesis Agent}
To investigate the effectiveness of the RL agent, we conduct an ablation study by introducing another agent that randomly selects LS operations for $T=10$ times (noted as w/o RL). The original setting with RL agent is noted as Ours. We test these two settings and summarized the results in Figure~\ref{fig:abs}.

We observe that although the agent with a random policy is capable of employing LS operations to simplify the circuit, it still falls short in terms of achieving optimal solving time reduction. To be specific, solving these 300 instances necessitates 7329,96 seconds for the  w/o RL and $6454.02$ seconds for our w/ RL agent, marking a reduction of $11.95\%$.

\subsubsection{Effectiveness of Cost-Customized Mapper}
To showcase the effectiveness of our proposed cost-customized mapper, we replace our cost-customized mapper with a mapper with conventional cost metrics. We denote the new setting as C. Mapper and our original setting as Ours. It is worth noting that both settings utilize the same RL agent with same parameters, and thus, we disregard the agent runtime ($\mathcal{T}_{agent}$) in the reported results. 

As shown in Figure~\ref{fig:abs}, the 300 instances processed by the conventional mapper take 9732.64 seconds to solve, which is 50.80\% longer than the solving time of Our Mapper. We attribute this difference to the divergent optimization objectives: the conventional mapper minimizes area and delay, while our cost-customized mapper targets solving complexity directly. 

\section{Conclusion and Future Work}\label{Sec:Conclusion}


This paper introduces a novel circuit-based preprocessing framework for circuit satisfiability problem, integrating logic synthesis with reinforcement learning to transform circuits into solver-friendly formats. Key innovations include an RL-guided logic synthesis pipeline for optimizing circuit transformations and a cost-customized LUT mapping strategy that minimizes branching complexity during SAT solving. Experiments on industrial benchmarks demonstrate significant runtime reductions—up to 63\% compared to baseline approaches—while maintaining compatibility with state-of-the-art solvers. Ablation studies validate the effectiveness of both the RL agent and cost-customized mapper in improving solving efficiency.

This work highlights the potential of combining EDA innovations with machine learning to advance SAT-solving methodologies. Beyond CSAT, the proposed framework offers a promising direction for tackling other SAT variants and related combinatorial optimization problems, enabling more efficient and scalable solutions. Future work will explore extending this approach to sequential circuits and broader EDA applications.

\balance

\section*{Acknowledgments}
This work was partly supported by the General Research Fund of the Hong Kong Research Grants Council (RGC) under Grant No. 14212422 and 14202824 and partly by the National Technology Innovation Center for EDA.

\newpage

\balance
\bibliographystyle{IEEEtran}
\bibliography{reference}

\end{document}